# Fully compensated and uncompensated ferrimagnetic ferrovalley semiconductors

Weifeng Xie[1], Libo Wang[1], Yunliang Yue[2], Xiong Xu[3], Huayan Xia[1], Hui Wang[3,*]

1 School of Microelectronics and Physics, Hunan University of Technology and Business, Changsha 410205, China

2 School of Information and Artificial Intelligence, Yangzhou University, Yangzhou 225127, China

3 School of Physics, Hunan Key Laboratory of Super Microstructure and Ultrafast Process, Hunan Key Laboratory of Nanophotonics and Devices, State Key Laboratory of Powder Metallurgy, Central South University, Changsha 410083, China

## Abstract

Altermagnets (AMs) and fully compensated ferrimagnets (FC-FIMs) are emerging classes of magnetic materials that combine the advantages of antiferromagnets and ferromagnets. Here, we elucidate the mechanism behind the uniaxial strain-driven transformation from AM to FC-FIM and find that the accompanying non-relativistic valley polarization is positively correlated with the net magnetic moment between magnetic atoms in opposite spin sublattices. We then propose an uncompensated ferrimagnetic monolayer VCrSeTeO to achieve large intrinsic valley polarization. Spin-orbit coupling (SOC) is shown to further increase the valley polarization to over 400 meV under uniaxial strains and the reason is explained in terms of SOC perturbation theorem. Furthermore, we reveal a distinctive anomalous valley Hall effect in which the valley Hall voltage is reversed within the same valley in ferrimagnet VCrSeTeO. This work proposes a strategy for realizing giant valley polarization and provides theoretical guidance for the application of ferrimagnetic ferrovalley semiconductors derived from altermagnets in valleytronics.

## 1. Introduction

Traditional collinear magnetic materials are categorized into ferromagnets (FM),

ferrimagnets (FIM), and antiferromagnets (AFM). Their key distinction lies in the magnitude and orientation of magnetic moments in different spin sublattices [1]. Recently, two specific classes of collinear magnetic materials—characterized by zero net magnetic moment yet opposite spins split in momentum space—have garnered widespread attention. These are termed two-dimensional altermagnets (AMs) [2,3] and fully compensated ferrimagnets (FC-FIM) [4,5], distinguishing them from conventional AFMs. A zero net magnetic moment (devoid of stray fields) endows AMs and FC-FIMs with robust immunity to external perturbations, ultrafast dynamic responses and ultrahigh spintronic storage densities as in AFMs [6–8]. Additionally, the lifted spin degeneracy in momentum space grants them the potential to generate pure spin currents, analogous to FMs or conventional FIMs [4,9]. In AMs and FC-FIMs, phenomena such as the anomalous Hall effect [10,11], Nernst effect [12,13], pure nonrelativistic spin-polarized currents [9], and magneto-optical Kerr effects [4,14,15] have been extensively investigated. Moreover, general methods for identifying AMs and FC-FIMs have been proposed [16,17], and the mutual transformation among AFM, AM, and FC-FIM via external stimuli (e.g., strain, external electric fields, and Janus engineering) has been elucidated [4,16,18–20].

In two-dimensional collinear magnetic materials, a rather prominent property is the co-utilization of the valley degree of freedom alongside the electron and spin degrees of freedom. In FMs, the term "ferrovalley" was first proposed to distinguish non-magnetic valley materials [21]. Unlike the valley splitting induced by spin-orbit coupling (SOC) in graphene [22,23] and non-magnetic transition metal dichalcogenides (e.g., $MoS_2$) [24], where an external magnetic field [25–27], circularly polarized light excitation [24,28,29], or magnetic proximity effect [30–32] is indispensable for the emergence of valley polarization (defined as $\Delta E_{c(v)} = E_{c(v)}(K) - E_{c(v)}(K')$). In ferrovalley materials, the intrinsic collinear magnetization is theoretically equivalent to an external magnetic field or magnetic proximity effect, inducing valley polarization spontaneously [21]. In AFMs, especially A-type van der Waals AFMs, external-field-induced valley polarization is associated with the layer degree of freedom [18,33–36], enabling flexible manipulation of the valley degree of freedom. It should be noted that whether in ferromagnetic ferrovalley or antiferromagnetic ferrovalley systems, SOC with out-of-plane magnetization needs to be considered for the generation of valley polarization [18,21,36].

In recently discovered AMs, their unique rotational and diagonal mirror symmetries enable uniaxial strain to break symmetry and induce nonrelativistic valley polarization—termed the piezovalley effect—without requiring SOC. Consequently, giant valley polarization can be achieved under sufficiently large uniaxial strain [9]. Subsequently, the concept of two-dimensional FC-FIMs was introduced, defined as magnetic materials with zero net magnetic moment but inequivalent local magnetic moments in opposite spin sublattices [4,5]. Under uniaxial strain, AMs can transform into FC-FIMs, giving rise to valley polarization [9,37]. In all aforementioned collinear magnetic materials, strain, external electric fields, or SOC are essential for generating valley polarization. We anticipate that magnetic materials exhibiting giant intrinsic valley polarization—without reliance on SOC or external fields—can be identified. Moreover, valley polarization in conventional FIMs remains underexplored, and it is unclear whether SOC plays a critical role in achieving giant valley polarization.

In this Letter, we reveal that valley polarization in uniaxial strain-modulated AMs is governed by the net magnetic moment between magnetic atoms in opposite spin sublattices, and we elucidate the mechanism underlying the AM-to-FC-FIM transformation. We then propose a strategy to achieve giant valley polarization: substituting magnetic atoms in one sublattice with those possessing different valence electrons, which enables giant intrinsic valley polarization in conventional FIMs. Moreover, SOC can further enhance valley polarization, and an anomalous valley Hall (AVH) effect is observed under out-of-plane magnetization. Leveraging the SOC perturbation theorem, we clarify the SOC-induced variation of valley polarization. Finally, a giant valley polarization exceeding 400 meV is achievable in conventional FIMs under moderate uniaxial strains and SOC.

## 2. Computational methods

Vienna *ab initio* simulation package (VASP) [38] based on density functional theory is used in the calculations, where the projector augmented wave method [39] and the Perdew-Burke-Ernzerhof (PBE) [40] exchange-correlation functional within the generalized gradient approximation are employed. A plane-wave cutoff energy of 560 eV is used consistently throughout the calculations. To eliminate interlayer interaction effects, a vacuum layer exceeding 15 Å is applied perpendicular to the two-dimensional plane. For structural optimization and self-consistent electronic structure calculations, the convergence criteria are set to $1 \times 10^{-7}$ eV for energy and 0.005 eV/Å for the force on each atom. A $25 \times 25 \times 1$ Γ-centered Monkhorst-Pack mesh of the *k*-point grid is used for the monolayers $V_2Se_2O$ and $V_2SeTeO$, while a $24 \times 25 \times 1$ *k*-point grid is employed for the monolayers $VCrSe_2O$ and $VCrSeTeO$. To treat the strong correlation effects of the *d* electrons in V and Cr atoms, the PBE + *U* method [41] is uniformly adopted, with effective *U* values of 4 eV for V [42] and 3.55 eV for Cr [43,44]. The Berry curvature and AVH conductivity are calculated using maximally localized Wannier functions as encoded in the WANNIER90 package [45].

## 3. Results and Discussion

Monolayers $V_2Se_2O$ and Janus $V_2SeTeO$ are identified as prototypical two-dimensional AMs. Their hallmark feature—uniaxial strain-driven nonrelativistic valley polarization (defined as $\Delta E_{c(v)} = E_{c(v)}(X) - E_{c(v)}(Y)$ )—renders these tetragonal monolayers as prominent piezovalley materials [9,42]. Top and side views of monolayers $V_2Se_2O$ and $V_2SeTeO$ are presented in Fig. 1 (a) and (b), respectively. The thick arrows with red and blue on V atoms in top view mark opposite spin sublattices of V, which presents zero net magnetic moment. We calculate the strain-dependent evolution of valley polarization in the uppermost valence band (UVB) under uniaxial strain along the *a*-axis for both $V_2Se_2O$ and $V_2SeTeO$, as shown by the red curves in Fig. 1(c) and (d), respectively. These results are in excellent agreement with available literature [9,42]. In contrast, we explicitly extract the net magnetic moment between the two V atoms within the unit cell under variational uniaxial strains along the *a*-axis, defined as $\Delta M(V_1 - V_2) = |M_{V_1}| - |M_{V_2}|$. It is found that the net magnetic moment $\Delta M(V_1 - V_2)$ varies quasi-linearly with uniaxial strain, indicating that uniaxial strain can induce a nonzero $\Delta M(V_1 - V_2)$ while preserving the zero total magnetic moment of monolayers $V_2Se_2O$ and $V_2SeTeO$, as confirmed by first-principles calculations. This phenomenon precisely corresponds to the defining characteristic of FC-FIMs [5,46,47]. As shown in Fig. S1, even under −5% compressive strain, monolayer $V_2Se_2O$ remains a semiconductor with an open bandgap and retains its zero net magnetic moment.

In contrast, for Janus monolayer $V_2SeTeO$, Figs. S2 and S3 reveal that −5% compressive strain closes the spin-down bandgap while leaving the spin-up bandgap open, resulting in a half-semimetal state. The equivalent integrated density of states (IDOS) for opposite spins near the Fermi level (Fig. S3) confirms that the total magnetic moment of $V_2SeTeO$ remains zero. Consequently, the transformation from AM to FC-FIM is realized under uniaxial strain [4]. Due to the quasi-linear relationship between $\Delta M(V_1 - V_2)$ and uniaxial strain, the trend of valley polarization versus $\Delta M(V_1 - V_2)$ is consistent with that versus uniaxial strain, as illustrated by the black lines in the insets of Fig. 1 (c) and (d). Evidently, a larger net magnetic moment between the two magnetic V atoms corresponds to larger valley polarization. In Fig. 1 (e) and (f), the band structures of $V_2Se_2O$ and $V_2SeTeO$ under −5% compressive strain are presented, respectively. For $V_2Se_2O$ ($V_2SeTeO$), at −5% compressive strain, the magnitude of $\Delta M(V_1 - V_2)$ is 0.012 $\mu_B$ (0.024 $\mu_B$), giving rise to the largest valley polarization of -85.20 meV (-167.41 meV) over the strain range from −5% compression to 5% tension. Intriguingly, −5% compressive strain drives $V_2SeTeO$ into a semimetallic state at the Y valley, analogous to strain-tuned monolayer $Nb_2SeTeO$ [48], thereby realizing the widely expected half-semimetallic FC-FIM [4]. This is verified by three key features: (i) zero total magnetic moment, (ii) inequivalent net magnetic moments between magnetic atoms, and (iii) equivalent IDOS for opposite spins around the Fermi level (shown in Fig. S3).

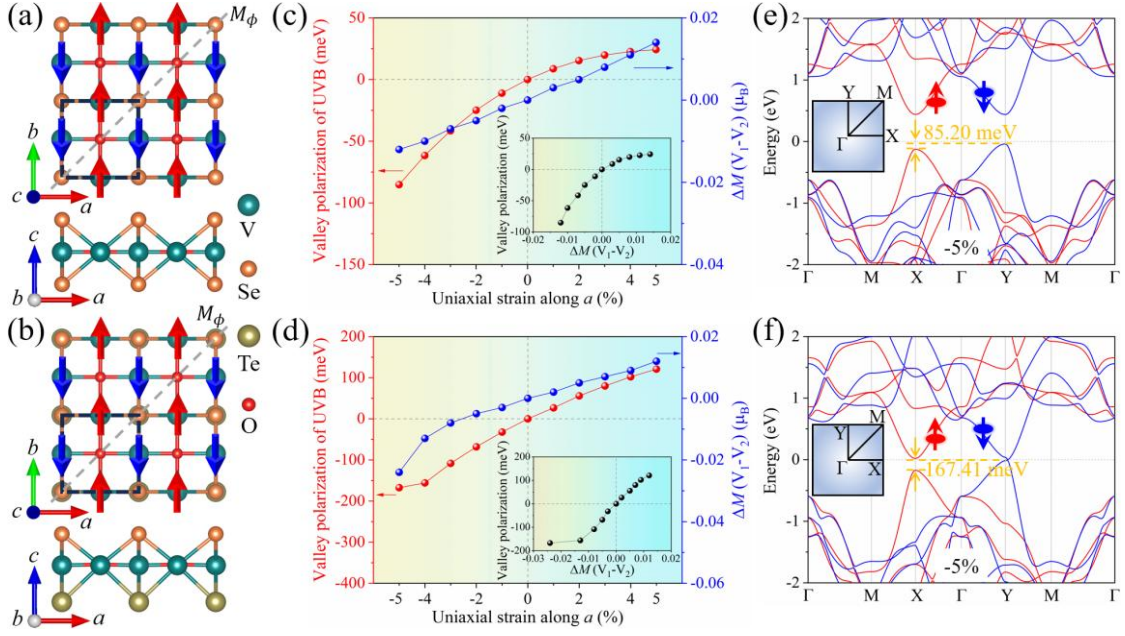

Fig. 1: Top and side views of monolayers (a) $V_2Se_2O$ and (b) $V_2SeTeO$, where the black dashed box represents the unit cell and the gray dashed line represents the diagonal mirror plane, thick red and blue arrows denote opposite local magnetic moments on the corresponding V atoms. Valley polarization of uppermost valence band (UVB) and net magnetic moment between V atoms in different spin sublattices ($\Delta M(V_1 - V_2)$) as functions of uniaxial strains along $a$ axis for (c) $V_2Se_2O$ and (d) $V_2SeTeO$, insets show the valley polarization versus the net magnetic moment $\Delta M(V_1 - V_2)$. Band structures of monolayers (e) $V_2Se_2O$ and (f) $V_2SeTeO$ under –5% compressive strain, where the red and blue lines represent spin-up and spin-down bands, respectively, the values of valley polarization on the UVB are marked, insets illustrate the high-symmetry path in the first Brillouin zone.

Elucidating the origin of the variational $\Delta M(V_1 - V_2)$ reveals the intrinsic mechanism underlying the uniaxial strain-driven transformation of AMs into FC-FIMs. Taking monolayer $V_2Se_2O$ as a prototype, we construct a V-centered octahedral intrinsic coordinate with $x$-$y$-$z$ axes to investigate the uniaxial strain-induced orbital hybridization between V and Se atoms or between V and O atoms. Figs. S4 and S5 display the $V_1$-centered and $V_2$-centered octahedral intrinsic coordinates, along with their corresponding projected density of states (PDOS), respectively. For spin-up $V_1$ atoms, we primarily focus on the hybridization of spin-up $d$-orbital electrons. The

PDOS distribution of the $d$-orbitals of $V_1$ and the $p$-orbitals of Se and O atoms is shown in Fig. S4 (b). It is found that the $d_{x^2-y^2}$ orbitals of $V_1$ exhibit dominant hybridization with the $p_x$ ($p_y$) orbitals of Se, while the hybridization with the $p$-orbitals of O near the Fermi level can be neglected. The conclusion derived from the PDOS analysis can also be corroborated by the hopping integration between the $d$-orbitals of $V_1$ and the $p$-orbitals of Se, as listed in Table SI. Evidently, only the $d_{x^2-y^2}$ orbitals of $V_1$ exhibit substantial hopping integration with the $p_x$ orbitals of $Se_{11}$ and $Se_{23}$ (labeled in Fig. S4 (a)) and the $p_y$ orbitals of $Se_{13}$ and $Se_{21}$ (also labeled in Fig. S4 (a)). Similarly, for spin-down $V_2$ atoms, based on the $V_2$-centered octahedral intrinsic coordinate (Fig. S5 (a)), only the hybridizations of spin-down $d$-orbital electrons with surrounding Se atoms are considered. The corresponding PDOS and hopping integration are presented in Fig. S5 (b) and Table SII, respectively. The results indicate that the $d_{x^2-y^2}$ orbitals of $V_2$ predominantly hybridize with the $p_x$ orbitals of $Se_{11}$ and $Se_{22}$ (labeled in Fig. S5(a)) as well as the $p_y$ orbitals of $Se_{12}$ and $Se_{21}$ (labeled in Fig. S5 (a)).

We concentrate on the variation of the total hopping integration between the $d_{x^2-y^2}$ orbitals of V and the $p_x$ ($p_y$) orbitals of surrounding Se atoms under uniaxial strain. The variational trend of hopping integration as a function of uniaxial strain along the $a$-axis is illustrated in Fig. 2. It is observed that the hopping integration between the $d_{x^2-y^2}$ orbitals of $V_2$ and the $p_x$ ($p_y$) orbitals of Se exhibits only a slight response to the change in uniaxial strain. In contrast, the hopping integration between the $d_{x^2-y^2}$ orbitals of $V_1$ and the $p_x$ ($p_y$) orbitals of Se shows a remarkable change under the action of uniaxial strain. Uniaxial compressive strain enhances the hopping integration of $V_1$, whereas it reduces the hopping integration of $V_2$ under uniaxial tensile strain. This distinct variational trend of hopping integration between the two V atoms leads to a non-zero $\Delta M(V_1 - V_2)$, which demonstrates an approximately linear dependence on strain. This mechanism clarifies the variation of $\Delta M(V_1 - V_2)$ with uniaxial strain in monolayer $V_2Se_2O$, as well as the primary origin of the transformation from AM to FC-FIM.

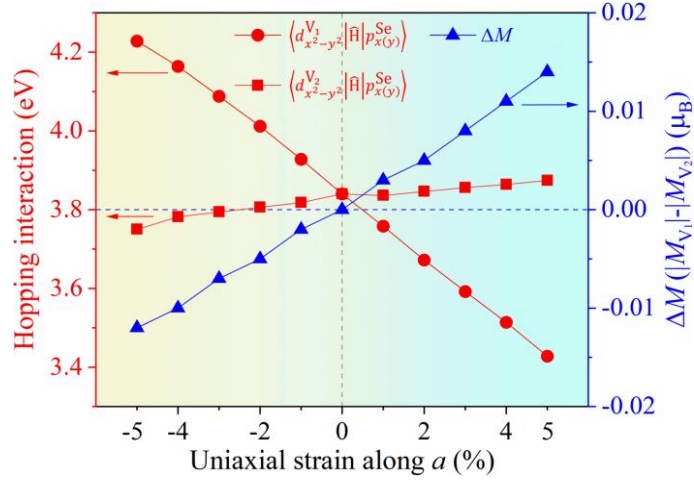

Fig. 2: Hopping integrations between $d$ orbitals of V atom ($V_1$, $V_2$) and $p$ orbitals of surrounding Se atoms as well as the net magnetic moment ($\Delta M(V_1 - V_2)$) as functions of uniaxial strains along $a$-axis in monolayer $V_2Se_2O$.

Fig. 1 (c) and (d) clarify the variational relationship among valley polarization, $\Delta M(V_1 - V_2)$ and uniaxial strain in monolayer $V_2Se_2O$ and $V_2SeTeO$, and reveal that valley polarization is positively correlated with $\Delta M$. We reasonably consider whether it is feasible to achieve a large $\Delta M$ by substituting one of the magnetic atoms in the spin sublattices with a transition metal atom of different valence electrons, thereby realizing giant valley polarization. Here, based on monolayer $V_2Se_2O$ and $V_2SeTeO$, we construct two stable monolayers $VCrSe_2O$ and $VCrSeTeO$ by substituting V with Cr. Fig. 3 (a) and (b) show the stable structures with opposite spins localized on V and Cr. The magnitudes of the local magnetic moments of V and Cr are distinct, as indicated by red and blue arrows with different lengths, characterizing the ferrimagnetic order in the monolayers $VCrSe_2O$ and $VCrSeTeO$. The dynamic, thermal, and structural stability of $VCrSe_2O$ has been investigated [49], the *ab initio* molecular dynamics (AIMD) simulation and phonon spectrum calculation of $VCrSeTeO$ are shown in Fig. S6 (a) and (b), respectively. The formation energy of $VCrSeTeO$ is also calculated in the SI. The results demonstrate that $VCrSeTeO$ exhibits thermodynamic, kinetic, and chemical stability. Besides, the morphology of different substituted sites with 50% Cr concentrations in a 2×2 supercell of $V_2SeTeO$ is schematically illustrated in Fig. S6 (c). In Fig. S6 (d), the relative energy of different substituted structures is calculated via first-principles calculations, which reveals that the P1 structure (corresponding to Fig. 3 (b)) is the most stable substituted configuration. After relaxation, the lattice constants of monolayers $VCrSe_2O$ and VCrSeTeO satisfy $a \neq b$. The diagonal mirror symmetry $M_\phi$, a key feature of AMs, is correspondingly broken (as shown in Fig. 3 (a) and (b)). From the perspective of symmetry, analogous to the diagonal mirror symmetry breaking induced by uniaxial strain in $V_2Se_2O$ and $V_2SeTeO$, a remarkable valley polarization can be obtained in monolayers $VCrSe_2O$ and VCrSeTeO without considering SOC.

Fig. 3(c) and (d) present the spin densities of monolayers $VCrSe_2O$ and VCrSeTeO, revealing that the signs of the spin densities are opposite and the magnitudes of the spin densities differ. The spin-down spin density is notably larger than the spin-up spin density, indicating a larger negative magnetic moment. First-principles calculations show that both monolayers $VCrSe_2O$ and VCrSeTeO have a total magnetic moment of -1 $\mu_B$, which is consistent with the distribution of spin density. Moreover, the FIM state is lower in energy than the FM state by 330.18 meV. In Fig. S7, we also illustrate the origin of the FIM state with a total magnetic moment of -1 $\mu_B$ in terms of the electron distribution on $d$-orbital splitting of crystal field. It is known that Cr has more valence electrons than V, which gives rise to a large $\Delta M$, leading to a predictable giant valley polarization. The band structures in Figs. 3 (e) and 3 (f) show intrinsic valley polarizations of 161.79 meV for $VCrSe_2O$ and 149.46 meV for VCrSeTeO. The non-zero $\Delta M(V_1 - V_2)$, non-zero total magnetic moment, and non-equivalent IDOS of opposite spins near the Fermi level confirm monolayer VCrSeTeO as conventional FIMs. For clarity, we hereafter refer to such systems as uncompensated FIMs (UFIMs) in contrast to FC-FIMs.

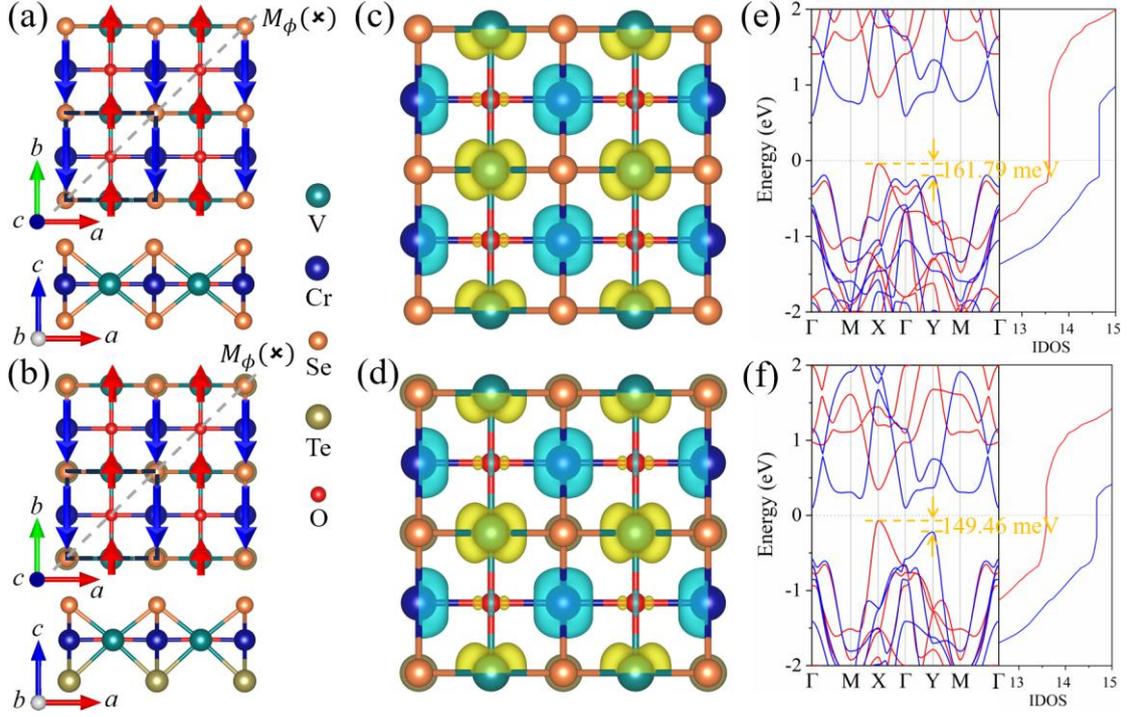

Fig. 3: Top and side views of monolayers (a) VCrSe$_2$O and (b) VCrSeTeO, where the black dashed box represents the unit cell and the gray dashed line represents the diagonal mirror plane (here the mirror symmetry along diagonal direction is broken), thick red and blue arrows denote opposite local magnetic moments on V and Cr atoms. Distributions of spin density of monolayers (c) VCrSe$_2$O and (d) VCrSeTeO, where the isosurface is set at $2 \times 10^{-2}$ $e$/Å$^3$ with yellow and cyan colors representing the accumulation of spin-up and spin-down electrons, respectively. Band structures and integrated density of states (IDOS) near the Fermi level for spin-up and spin-down electrons in monolayers (e) VCrSe$_2$O and (f) VCrSeTeO, where red and blue lines represent spin-up and spin-down, respectively, the values of valley polarization of the UVB are marked on the band structures.

Comparing the bands in Fig. 3 (e) with those in Fig. 3 (f), there is no interference from other energy bands within the energy interval between the X and Y valleys in UFIM VCrSeTeO, which facilitates the detection and utilization of valley polarization. Thus, we prioritize exploring valley polarization of VCrSeTeO. In conventional ferrovalley materials [21,50], SOC is the essential origin of valley polarization. For UFIM VCrSeTeO, we incorporate SOC and present the corresponding band structures with magnetizations along three principal axes in Fig. 4 (a)-(c). Notably, the band with [010] magnetization exhibits a larger valley polarization (193.02 meV) than the intrinsic state (149.46 meV). Smaller valley polarization is observed for the band with [100] magnetization, while the band with [001] magnetization shows nearly unchanged valley polarization. This phenomenon demonstrates that SOC also exerts a significant impact on valley polarization when magnetizations are oriented along [010] and [100]. Especially, [010] magnetization enhances the valley polarization by nearly 30%, yielding a valley polarization of almost 200 meV.

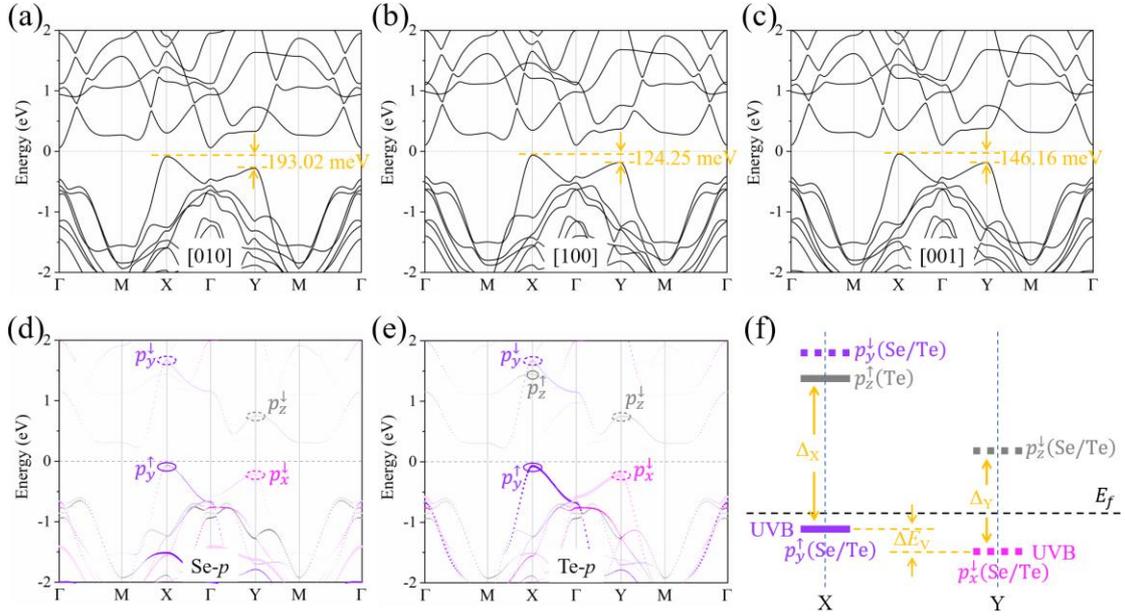

Fig. 4: Band structures of VCrSeTeO under SOC with (a) [010], (b) [100] and (c) [001] magnetizations, where the values of valley polarization on the UVB are marked. Distribution of $p$ orbitals of Se and Te near the Fermi level on the band structures without SOC. (f) Schematic illustration of the orbital projections near the Fermi level at X and Y valleys, where the valley polarization ($\Delta E_v$) and corresponding bandgaps ($\Delta_X$ and $\Delta_Y$) are marked.

Herein, we employ the SOC perturbation theorem to elucidate the origin of distinct valley polarizations under different magnetizations. By analyzing the atom-resolved band structures (as depicted in Fig. 4(d), (e), and Fig. S9), we find that the $p$-orbitals of Se and Te make a dominant contribution to the UVB at the X and Y valleys. Notably, the $p_y^\uparrow$ and $p_x^\downarrow$ orbitals of Te predominantly contribute to the X and Y valleys around the Fermi energy, respectively. In contrast, the contribution from the $d$-orbitals of V and Cr to the X and Y valleys is almost negligible. Fig. 4 (f) illustrates the schematic of orbitals near the Fermi level, corresponding to Fig. 4 (d) and (e). The valley polarization ($\Delta E_v$) in monolayer VCrSeTeO is defined as the energy difference between the UVB at the X and Y valleys, i.e., $\Delta E_v = E_v(X) - E_v(Y)$. The SOC Hamiltonian can be written as [51] $\hat{H} = \hat{H}_{SOC} + \hat{H}'_{SOC}$, $\hat{H}_{SOC} = \lambda \hat{S}_{z'} \left( \hat{L}_z \cos\theta + \frac{1}{2}\hat{L}_+ e^{-i\varphi} \sin\theta + \frac{1}{2}\hat{L}_- e^{i\varphi} \sin\theta \right)$, $\hat{H}'_{SOC} = \frac{\lambda}{2} \hat{S}_{+'} \left( -\hat{L}_z \sin\theta - \hat{L}_+ e^{-i\varphi} \sin^2\frac{\theta}{2} + \hat{L}_- e^{i\varphi} \cos^2\frac{\theta}{2} \right) + \frac{\lambda}{2} \hat{S}_{-'} \left( -\hat{L}_z \sin\theta + \hat{L}_+ e^{-i\varphi} \cos^2\frac{\theta}{2} - \hat{L}_- e^{i\varphi} \sin^2\frac{\theta}{2} \right)$, where $\hat{H}_{SOC}$ and $\hat{H}'_{SOC}$ are spin-conserving and spin-non-conserving SOC Hamiltonian, respectively, $\lambda$ represents the SOC coefficient, $\hat{S}_{z'}$ and $\hat{L}_z$ represent the $z'$ or $z$ components of spin and orbital angular momentum, respectively, $\theta$ and $\varphi$ are the polar angle and azimuthal angle of the spin, respectively, and the ladder operators are given by $\hat{L}_\pm = \hat{L}_x \pm i\hat{L}_y$ and $\hat{S}_\pm = \hat{S}_x \pm i\hat{S}_y$. Table I summarizes the coupling relationships between different $p$-orbitals under the action of $\hat{H}_{SOC}$. Based on this, when the magnetization is along the [010] direction (i.e., $\theta = 90°$ and $\varphi = 90°$), we find that $\langle p_y^\uparrow | \hat{H}_{SOC} | p_z^\uparrow \rangle = 0$ and $\langle p_y^\uparrow | \hat{H}'_{SOC} | p_y^\downarrow \rangle = 0$ at the X valley, leaving the bandgap at the X valley unchanged. However, the bandgap at the Y valley is enlarged by $\frac{2\lambda^2}{\Delta_0}$ due to $\langle p_x^\downarrow | \hat{H}_{SOC} | p_z^\downarrow \rangle = i\lambda$ (from Table I). Consequently, the unchanged bandgap at the X valley and the enlarged bandgap at the Y valley contribute to the increase of $\Delta E_v$ when the magnetization is along [010], with an enhancement rate of nearly 30% compared to the intrinsic state. Similarly, when the magnetization is along [100] (i.e., $\theta = 90°$ and $\varphi = 0°$), Table I shows that $\langle p_y^\uparrow | \hat{H}_{SOC} | p_z^\uparrow \rangle = -i\lambda$ and $\langle p_y^\uparrow | \hat{H}'_{SOC} | p_y^\downarrow \rangle = 0$ at the X valley, while $\langle p_x^\downarrow | \hat{H}_{SOC} | p_z^\downarrow \rangle = 0$ at the Y valley. This leads to an enlarged bandgap of $\frac{2\lambda^2}{\Delta_0}$ at the X valley and an unchanged bandgap at the Y valley, which reduces $\Delta E_v$ with a decrement rate of 17%. Moreover, comparing the bandgaps $\Delta_X$ and $\Delta_Y$ (shown in Fig. 4(f)) reveals $\Delta_X > \Delta_Y$. According to enlarged bandgap value $\frac{2\lambda^2}{\Delta_0}$, the smaller bandgap $\Delta_0$ without SOC perturbation, the larger increment of bandgap under the SOC perturbation. It is reasonable that the increased rate of valley polarization under [010] magnetization is larger than the decreased rate under [100] magnetization. Finally, the nearly unchanged valley polarization under [001] magnetization ($\theta = 0°$) arises from the coupling relations $\langle p_y^\uparrow|\hat{H}_{SOC}|p_z^\uparrow\rangle = 0$, $\langle p_y^\uparrow|\hat{H}'_{SOC}|p_y^\downarrow\rangle = 0$ and $\langle p_x^\downarrow|\hat{H}_{SOC}|p_z^\downarrow\rangle = 0$ as tabulated in Table I.

Table I: Coupling strengths between different *p*-orbitals via the spin-conserving SOC Hamiltonian.

| Orbitals | $p_x$ | $p_y$ | $p_z$ |
|---|---|---|---|
| $p_x$ | 0 | $-i\lambda\cos\theta$ | $i\lambda\sin\theta\sin\varphi$ |
| $p_y$ | $i\lambda\cos\theta$ | 0 | $-i\lambda\sin\theta\cos\varphi$ |
| $p_z$ | $-i\lambda\sin\theta\sin\varphi$ | $i\lambda\sin\theta\cos\varphi$ | 0 |

In conventional ferrovalley materials, the AVH effect can be induced when SOC with [001] magnetization is taken into account [21,50]. Here, we schematically illustrate the AVH effect existing in UFIM VCrSeTeO. The three-dimensional energy distribution of magnetocrystalline anisotropy energy (MAE) of UFIM VCrSeTeO is presented in Fig. 5 (a), where the magnetic easy and hard axes are along [010] and [100] directions, respectively. The MAE is expressed as $E_{[001]} - E_{[010]} = 463.08$ μeV, indicating that an external magnetic field can readily switch the magnetization from the magnetic easy axis ([010] direction) to [001] and [00$\bar{1}$] directions. The band structures and the distribution of Berry curvature at high-symmetry points for magnetizations along [001] and [00$\bar{1}$] directions are calculated and shown in Fig. 5 (b). It is found that the Berry curvatures at the X valley are reversed when the magnetization is reversed from [001] to [00$\bar{1}$]. The finite Berry curvature at the X valley can induce in-plane AVH conductivity. Fig. 5 (c) shows the AVH conductivity within the energy range spanned by the X and Y valleys is non-zero and opposite for magnetizations along [001] and [00$\bar{1}$] directions, respectively. A schematic of the AVH effect is depicted in Fig. 5 (d). Unlike the conventional AVH effect in ferromagnetic ferrovalley materials [21], when moderate hole doping and an external electric field along the *x*-axis are simultaneously applied, opposite AVH voltages along the *y*-axis emerge at the same valley (i.e., the X valley in the ferrimagnetic ferrovalley VCrSeTeO). The underlying mechanism is that the opposite non-zero Berry curvature along *z*-axis only at X valley as a pseudo magnetic field derives holes to deflect along *y*-axis, thus generating opposite AVH voltages at the X valley.

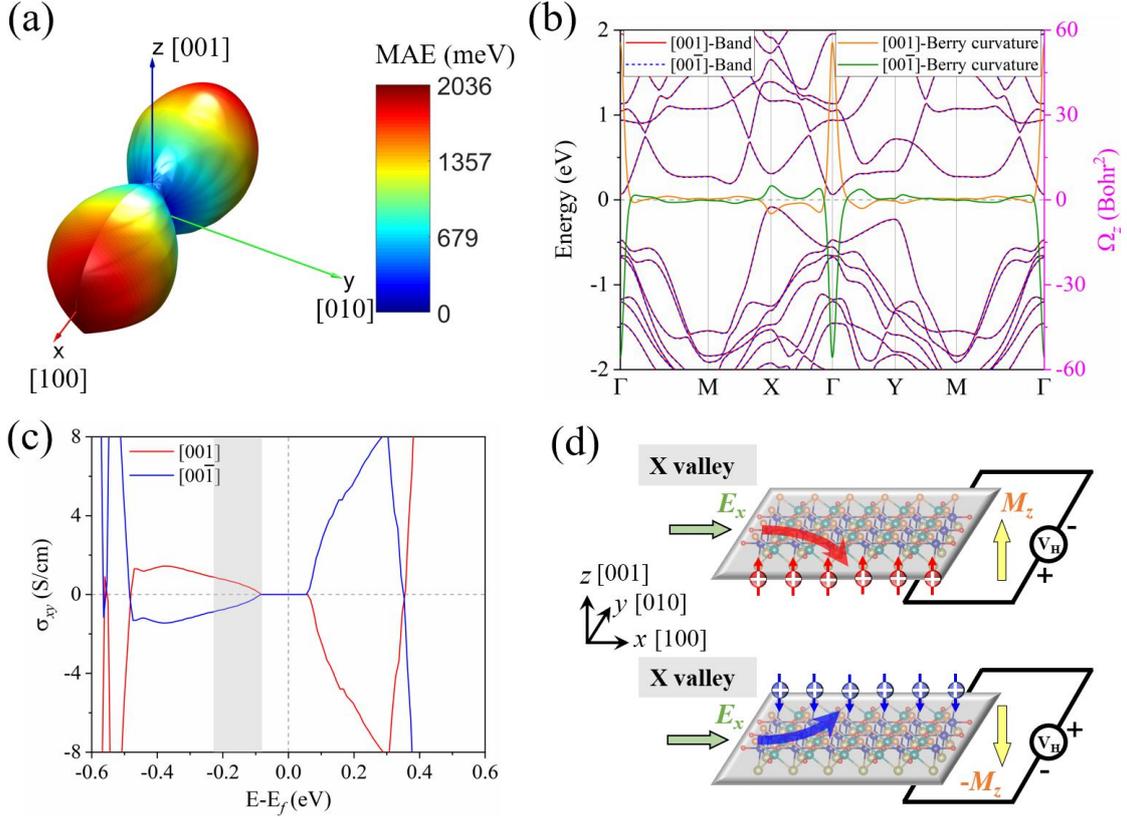

Fig. 5: (a) Three-dimensional distribution of the magnetocrystalline anisotropy energy (MAE) in VCrSeTeO. (b) Band structures considering SOC with [001] and [00$\bar{1}$] magnetizations, and the distribution of Berry curvature along high-symmetry paths. (c) Anomalous valley Hall conductivity ($\sigma_{xy}$) as a function of energy for the [001] and [00$\bar{1}$] magnetizations, the gray region indicates the energy range between X and Y valleys. (d) Schematic illustration of the anomalous valley Hall effect in the conventional FIM VCrSeTeO.

Similar to AMs, uniaxial strain can induce valley polarization and enhance it with increasing uniaxial strain. For UFIM VCrSeTeO, 5% tensile uniaxial strain increases the nonrelativistic valley polarization to 264.04 meV. When SOC with [010] magnetization is considered, the valley polarization reaches 310.67 meV. Fig. S9 (a) shows the variation of valley polarization with uniaxial strain along the *a*-axis (from −5% to 5%) with and without SOC. It is evident that SOC with [010] magnetization universally enhances the nonrelativistic valley polarization, and the underlying mechanism aligns with that of the intrinsic case (as explained via the SOC perturbation theorem). In Fig. S9 (b), uniaxial strain along the *b*-axis also modulates the valley polarization. A −5% compressive uniaxial strain along the *b*-axis increases the nonrelativistic valley polarization to 385.65 meV, and SOC with [010] magnetization further increase it to a giant value of 414.15 meV. Additionally, the same trend where SOC with [010] magnetization increases valley polarization with *b*-axis uniaxial strain persists. Overall, the realization of giant valley polarization in UFIM VCrSeTeO is primarily attributed to the large net magnetic moment between magnetic atoms in opposite spin sublattices after substituting V in $V_2$SeTeO with Cr, and uniaxial strains and SOC with [010] magnetization can further increase the valley polarization significantly.

## 4. Conclusion

In summary, first-principles calculations reveal that the valley polarization induced by uniaxial strain in two-dimensional AMs correlates with the net magnetic moment between magnetic atoms belonging to opposite spin sublattices of FC-FIM. Based on this correlation, we propose a strategy for achieving giant valley polarization: substituting one magnetic atom in AMs with another magnetic atom that has a different number of valence electrons. It is found that the Cr-substituted ferrimagnetic semiconductor VCrSeTeO, formed by replacing one V atom in AM $V_2$SeTeO, exhibits an intrinsic giant valley polarization. Moreover, the SOC with [010] magnetization further significantly enhances the valley polarization. In terms of SOC perturbation theorem, we elucidate the origin of the valley-polarization dependence on the magnetization direction. We also demonstrate that an out-of-plane magnetization in UFIM VCrSeTeO can give rise to an AVH effect, which is distinct from that in conventional ferromagnetic ferrovalley materials. Furthermore, under the consideration of both uniaxial strain and SOC, a giant valley polarization exceeding 400 meV is obtained in UFIM VCrSeTeO. The results suggest that UFIM derived from AMs hold promising potential for applications in valleytronics.

5. Acknowledgments

This work was supported by the National Natural Science Foundation of China (Grants No. 12504071), the Natural Science Foundation of Hunan Province (Grant No. 2026JJ60129), the Changsha Natural Science Foundation (Grant No. kg2502211), the National College Student Innovation Training Program Project (Grant No. 202510554047), the Postdoctoral Fellowship Program of the China Postdoctoral Science Fund (Grant No. GZC20252247).

Reference

[1] B. D. Cullity and C. D. Graham, *Introduction to Magnetic Materials* (John Wiley & Sons, 2011).

[2] L. Šmejkal, J. Sinova, and T. Jungwirth, Beyond Conventional Ferromagnetism and Antiferromagnetism: A Phase with Nonrelativistic Spin and Crystal Rotation Symmetry, Phys. Rev. X **12**, 031042 (2022).

[3] L. Šmejkal, J. Sinova, and T. Jungwirth, Emerging Research Landscape of Altermagnetism, Phys. Rev. X **12**, 040501 (2022).

[4] Y. Liu, S.-D. Guo, Y. Li, and C.-C. Liu, Two-Dimensional Fully Compensated Ferrimagnetism, Phys. Rev. Lett. **134**, 116703 (2025).

[5] S.-D. Guo, Q. Luo, S.-H. Zhang, and P. Jiang, External field induced transition from altermagnetic metal to fully compensated ferrimagnetic metal in monolayer Cr 2 O, Phys. Rev. B **113**, 064408 (2026).

[6] V. Baltz, A. Manchon, M. Tsoi, T. Moriyama, T. Ono, and Y. Tserkovnyak, Antiferromagnetic spintronics, Rev. Mod. Phys. **90**, 1 (2018).

[7] T. Jungwirth, X. Marti, P. Wadley, and J. Wunderlich, Antiferromagnetic spintronics, Nature Nanotech **11**, 3 (2016).

[8] T. Jungwirth, J. Sinova, A. Manchon, X. Marti, J. Wunderlich, and C. Felser, The

multiple directions of antiferromagnetic spintronics, Nature Phys **14**, 3 (2018).

[9] H.-Y. Ma, M. Hu, N. Li, J. Liu, W. Yao, J.-F. Jia, and J. Liu, Multifunctional antiferromagnetic materials with giant piezomagnetism and noncollinear spin current, Nat. Commun. **12**, 2846 (2021).

[10] L. Šmejkal, R. González-Hernández, T. Jungwirth, and J. Sinova, Crystal time-reversal symmetry breaking and spontaneous Hall effect in collinear antiferromagnets, Sci. Adv. **6**, eaaz8809 (2020).

[11] L. Zhang, S.-D. Guo, and G. Zhu, Electric-field-induced fully compensated ferrimagnetism in experimentally synthesized monolayer MnSe, Applied Physics Letters **127**, 142405 (2025).

[12] Y. Yang et al., Altermagnet-Driven Magnon Spin Splitting Nernst Effect, Phys. Rev. Lett. **136**, 026701 (2026).

[13] A. Badura et al., Observation of the anomalous Nernst effect in altermagnetic candidate Mn5Si3, Nat Commun **16**, 7111 (2025).

[14] M. Su, D. Zhang, H. Ye, G. P. Zhang, M. Gu, and J. Wang, Interlayer-sliding controlled magneto-optical effect and ferrovalley in a fully compensated ferrimagnetic bilayer, Phys. Rev. B **112**, 195427 (2025).

[15] L. Bai, W. Feng, S. Liu, L. Šmejkal, Y. Mokrousov, and Y. Yao, Altermagnetism: Exploring New Frontiers in Magnetism and Spintronics, Adv. Funct. Mater. (2024).

[16] S.-D. Guo, S. Chen, and G. Wang, Spin ordering induced fully compensated ferrimagnetism achieved in bilayers of Cr 2 C 2 S 6, Phys. Rev. B **112**, 134430 (2025).

[17] S.-D. Guo, J. He, and Y. S. Ang, Achieving fully compensated ferrimagnetism through two-dimensional CrI3/CrGeTe3 heterojunctions, Applied Physics Letters **127**, 232401 (2025).

[18] S.-D. Guo, W. Xu, Y. Xue, G. Zhu, and Y. S. Ang, Layer-locked anomalous valley Hall effect in a two-dimensional A -type tetragonal antiferromagnetic insulator, Phys. Rev. B **109**, 134426 (2024).

[19] S.-D. Guo, Y.-L. Tao, Z.-Y. Zhuo, G. Zhu, and Y. S. Ang, Electric-field-tuned

anomalous valley Hall effect in A -type hexagonal antiferromagnetic monolayers, Phys. Rev. B **109**, 134402 (2024).

[20] S.-D. Guo, Valley polarization in two-dimensional zero-net-magnetization magnets, Appl. Phys. Lett. **126**, 080502 (2025).

[21] W.-Y. Tong, S.-J. Gong, X. Wan, and C.-G. Duan, Concepts of ferrovalley material and anomalous valley Hall effect, Nat Commun **7**, 1 (2016).

[22] D. Xiao, W. Yao, and Q. Niu, Valley-Contrasting Physics in Graphene: Magnetic Moment and Topological Transport, Phys. Rev. Lett. **99**, 236809 (2007).

[23] A. Rycerz, J. Tworzydło, and C. W. J. Beenakker, Valley filter and valley valve in graphene, Nature Phys **3**, 172 (2007).

[24] K. F. Mak, K. He, J. Shan, and T. F. Heinz, Control of valley polarization in monolayer MoS2 by optical helicity, Nature Nanotech **7**, 494 (2012).

[25] Y. Li et al., Valley Splitting and Polarization by the Zeeman Effect in Monolayer MoSe 2, Phys. Rev. Lett. **113**, 266804 (2014).

[26] D. MacNeill, C. Heikes, K. F. Mak, Z. Anderson, A. Kormányos, V. Zólyomi, J. Park, and D. C. Ralph, Breaking of Valley Degeneracy by Magnetic Field in Monolayer MoSe 2, Phys. Rev. Lett. **114**, 037401 (2015).

[27] G. Aivazian, Z. Gong, A. M. Jones, R.-L. Chu, J. Yan, D. G. Mandrus, C. Zhang, D. Cobden, W. Yao, and X. Xu, Magnetic control of valley pseudospin in monolayer WSe2, Nature Phys **11**, 148 (2015).

[28] K. F. Mak, K. L. McGill, J. Park, and P. L. McEuen, The valley Hall effect in MoS 2 transistors, Science **344**, 1489 (2014).

[29] W.-T. Hsu, Optically initialized robust valley-polarized holes in monolayer WSe2, NATURE COMMUNICATIONS (2015).

[30] H. Matsuoka, T. Habe, Y. Iwasa, M. Koshino, and M. Nakano, Spontaneous spin-valley polarization in NbSe2 at a van der Waals interface, Nat Commun **13**, 5129 (2022).

[31] D. Zhong et al., Layer-resolved magnetic proximity effect in van der Waals heterostructures, Nat. Nanotechnol. **15**, 187 (2020).

[32] M. Abdollahi and M. B. Tagani, Tuning the magnetic properties of a VSe 2

monolayer via the magnetic proximity effect mediated by Zeeman-type spin-orbit interaction, Phys. Rev. B **108**, 024427 (2023).

[33] A. Gao et al., Layer Hall effect in a 2D topological axion antiferromagnet, Nature **595**, 521 (2021).

[34] X. Li, T. Cao, Q. Niu, J. Shi, and J. Feng, Coupling the valley degree of freedom to antiferromagnetic order, Proc. Natl. Acad. Sci. U.S.A. **110**, 3738 (2013).

[35] T. Zhang, X. Xu, B. Huang, Y. Dai, L. Kou, and Y. Ma, Layer-polarized anomalous Hall effects in valleytronic van der Waals bilayers, Mater. Horiz. **10**, 483 (2023).

[36] Y. Liu, J. Dong, G. Ni, and G. Gao, Universal strategy for spin and valley control: Electric-field-induced splitting and anomalous valley Hall effect in antiferromagnetic bilayers, Phys. Rev. B **113**, 014422 (2026).

[37] Y. Huang, C. Hua, R. Xu, J. Liu, Y. Zheng, and Y. Lu, Spin Inversion Enforced by Crystal Symmetry in Ferroelastic Altermagnets, Phys. Rev. Lett. **135**, 266701 (2025).

[38] G. Kresse and J. Furthmüller, Efficiency of ab-initio total energy calculations for metals and semiconductors using a plane-wave basis set, Computational Materials Science **6**, 1 (1996).

[39] P. E. Blöchl, Projector augmented-wave method, Phys. Rev. B **50**, 24 (1994).

[40] J. P. Perdew, K. Burke, and M. Ernzerhof, Generalized Gradient Approximation Made Simple, Phys. Rev. Lett. **77**, 18 (1996).

[41] S. L. Dudarev, G. A. Botton, S. Y. Savrasov, C. J. Humphreys, and A. P. Sutton, Electron-energy-loss spectra and the structural stability of nickel oxide: An LSDA+U study, Phys. Rev. B **57**, 3 (1998).

[42] Y. Zhu, T. Chen, Y. Li, L. Qiao, X. Ma, C. Liu, T. Hu, H. Gao, and W. Ren, Multipiezo Effect in Altermagnetic $V_2$SeTeO Monolayer, Nano Lett. **24**, 472 (2024).

[43] X. Chen, D. Wang, L. Li, and B. Sanyal, Giant spin-splitting and tunable spin-momentum locked transport in room temperature collinear antiferromagnetic semimetallic CrO monolayer, Appl. Phys. Lett. **123**, 022402 (2023).

[44] S.-D. Guo, X.-S. Guo, K. Cheng, K. Wang, and Y. S. Ang, Piezoelectric altermagnetism and spin-valley polarization in Janus monolayer Cr2SO, Appl. Phys. Lett. **123**, 082401 (2023).

[45] N. Marzari, A. A. Mostofi, J. R. Yates, I. Souza, and D. Vanderbilt, Maximally localized Wannier functions: Theory and applications, Rev. Mod. Phys. **84**, 1419 (2012).

[46] S.-D. Guo and Y. S. Ang, Spontaneous spin splitting in electric potential difference antiferromagnetism, Phys. Rev. B **108**, L180403 (2023).

[47] S.-D. Guo, Y.-L. Tao, G. Wang, and Y. S. Ang, How to produce spin-splitting in antiferromagnetic materials, J. Phys.: Condens. Matter **36**, 215804 (2024).

[48] W. Xie, X. Xu, Y. Yue, H. Xia, and H. Wang, Piezovalley effect and magnetovalley coupling in altermagnetic semiconductors studied by first-principles calculations, Phys. Rev. B **111**, 134429 (2025).

[49] W. Xie, L. Wang, X. Xu, Y. Yue, H. Xia, L. He, H. Wang, School of Microelectronics and Physics, Hunan University of Technology and Business, Changsha 410205, China, School of Physics, Central South University, Changsha 410083, China, and College of Information Engineering, Yangzhou University, Yangzhou 225127, China, Realizing giant valley polarization effect based on monolayer altermagnets, Acta Phys. Sin. **74**, 227502 (2025).

[50] W. Xie, L. Zhang, Y. Yue, M. Li, and H. Wang, Giant valley polarization and perpendicular magnetocrystalline anisotropy energy in monolayer M X 2 ( M = Ru , Os ; X = Cl , Br ), Phys. Rev. B **109**, 024406 (2024).

[51] H. Xiang, C. Lee, H.-J. Koo, X. Gong, and M.-H. Whangbo, Magnetic properties and energy-mapping analysis, Dalton Trans. **42**, 4 (2013).